\documentclass[11pt,preprint]{aastex}

\newcommand{\etal}{{et al.}}
\newcommand{\eg}{{e.g.,}}
\newcommand{\ie}{{i.e.,}}
\newcommand{\kms}{km~s$^{-1}$~}
\newcommand{\txs}{TXS\,0145+386}
\newcommand{\fourc}{4C\,15.55}

\begin{document}

\title{Extremely Red Objects in Two Quasar Fields at z\,$\sim$\,1.4\footnotemark[1]}

\footnotetext[1]{Based in part on data collected 
at the Subaru Telescope, which is operated by the National Astronomical Observatory 
of Japan; on data obtained with the Chandra X-Ray Observatory, which is 
operated for the National Aeronautics and Space Administration by the Smithsonian
Astrophysical Observatory; on data obtained at the Canada-France-Hawaii 
Telescope, which is operated by the National Research Council of 
Canada, the Institut National des Sciences de l'Univers of the Centre National de 
la Recherche Scientifique of France, and the University of Hawaii; and on data 
obtained at the The United Kingdom Infrared 
Telescope, which is operated by the Joint Astronomy Centre on behalf of the U.K.
Particle Physics and Astronomy Research Council. Some of the data presented herein 
were obtained at the W. M. Keck 
Observatory, which is operated as a scientific partnership among the California 
Institute of Technology, the University of California and the National Aeronautics 
and Space Administration. The Observatory was made possible by the
financial support of the W. M. Keck Foundation.}

\author{Alan Stockton\altaffilmark{2} and Elizabeth McGrath}
\affil{Institute for Astronomy, University of Hawaii, 2680 Woodlawn
 Drive, Honolulu, HI 96822}

\and

\author{Gabriela Canalizo\altaffilmark{2}}
\affil{Institute of Geophysics and Planetary Physics and
Department of Physics, University of California, Riverside, CA 92521}

\altaffiltext{2}{Visiting Astronomer at the Infrared Telescope Facility, which is 
operated by the University of Hawaii under Cooperative Agreement no. 
NCC 5-538 with the National Aeronautics and Space Administration, Office of 
Space Science, Planetary Astronomy Program.}

\begin{abstract}

We present an investigation of the properties and environments of bright extremely
red objects (EROs) found in the fields of the quasars \txs\ and \fourc,
both at $z\sim1.4$.  
There is marginal evidence from {\it Chandra} ACIS imaging for hot cluster
gas with a luminosity of a few $10^{44}$ ergs s$^{-1}$ in the field of \fourc. 
The \txs\ field has an upper limit at a similar value, but it also clearly shows 
an overdensity of faint galaxies. None of the EROs are detected as X-ray
sources.  For two of the EROs that have spectral-energy distributions and
rest-frame near-UV spectra that show that they are strongly dominated by old stellar
populations, we determine radial-surface-brightness profiles from
adaptive-optics images.  Both of these galaxies are best fit by profiles close to
exponentials, plus a compact nucleus comprising $\sim30$\% of the
total light in one case and 8\% in the other.  Neither is well fit by an $r^{1/4}$-law
profile.  This apparent evidence for the formation of massive $\sim2\times10^{11}$
disks of old stars in the early universe indicates that at least some galaxies
formed essentially monolithically, with high star-formation rates sustained over
a few $10^8$ years, and without the aid of major mergers.

\end{abstract}


\section{Introduction}

Galaxies classified as
{\it extremely red objects} (EROs; \eg\ \citealt{dad00,cha00,sto01,cim02,sar04,yan04,
dad04}; earlier references can be found in these papers)
are usually defined to have $R\!-\!K$ greater than a value somewhere in
a range of 5--6. They have long been known to include at least
two disparate classes.  Some are highly reddened high-redshift star-forming galaxies 
(\eg\ \citealt{dey99,cim02,yan04b}).  The remainder are almost exclusively galaxies
with $z\gtrsim1$ whose stellar populations are already
ancient and have little admixture from more recent star formation
\citep[\eg][]{sto95,dun96,spi97,soi99,liu00,sto01,sto04,che04,cim04,fu05}. 
For convenience, we shall 
henceforth simply refer to this latter class as ``old galaxies'' (OGs;
however, note that
this term refers to the age of the stellar population, not necessarily 
the age of assembly of the galaxy itself).
These OGs represent the earliest major episodes of star
formation in the universe, and they seem to have formed the
overwhelming majority of their stars over a very short period. 
Accordingly, they potentially 
offer us an avenue to investigate the formation and early evolution of the
very first massive galaxies to appear in the universe.

There have been a number of discussions of the mix of morphologies in
various samples of EROs.  It is commonly assumed that morphological types are
surrogates for the two classes of EROs mentioned above; \ie\ that spheroidals as
a group
are more-or-less congruent with galaxies dominated by old stellar populations
and that disk-like galaxies and those showing obvious merging activity
are dusty star-forming galaxies \citep{mor00,sti01,smi02,saw05}.  Nevertheless, 
both \citet{gil03} and
\citet{yan04} mention the possibility of truly passive disk-dominated galaxies.  
\citet{iye03} seem to have found such a galaxy at $z\sim1.5$, and
\citet{sto04} and \citet{fu05} have shown examples of galaxies that seem to be
essentially pure disks of old stars at $z=2.48$ and $z=1.34$, respectively. 

So far, the morphologies of only a few OGs at redshifts significantly greater than 1
have been studied in any detail (\eg\ \citealt{bun03,sto04,yan04,cim04,fu05,dad05}). 
Here we present an investigation of
the morphologies and environments of 3 OGs (and 1 dusty star-forming galaxy) 
found in the fields of the quasars TXS\,0145+386 ($z=1.442$) and 4C\,15.55
($z=1.406$). We assume a flat cosmology with $H_0 = 70$ and $\Omega_m=0.3$.

\section{Observations and Data Reduction}

\subsection{Optical and Near-IR Observations}

The two fields discussed in this paper are part of a larger sample
identified from a survey of the fields of (radio-loud) quasars with $1.4<z<1.6$.
We have chosen to observe quasar fields in this redshift range for three reasons. First, there is
the simple and practical observational matter that, for our near-IR observations,
it is necessary to have one or more objects in each field
that will be detected in single short exposures and that can be used to center
each dithered frame that is coadded to make the final image. Centering the field
on a quasar ensures that at least one such object will be present, even when
our field size is relatively small. Second, radio sources at high redshifts appear 
to mark regions that include 
those of high overdensity (\eg\ \citealt{bes00,bes03,bar03,rot04}), which are precisely the 
regions in which most models expect
processes of galaxy formation and evolution to have proceeded most rapidly.
Significantly, compelling results from the 2dF survey show that radio quiet QSOs are
much less strongly clustered than are quasars (\citealt{cro01}). 
Finally, and most importantly, 
looking for companions to radio sources {\it at a specific redshift} allows us to choose redshifts for
which the photometric diagnostics from standard broadband filters give the cleanest separation
between old galaxy populations and highly reddened star-forming galaxies or other 
possible contaminants.
Specifically, for $z\sim1.5$, the 4000 \AA\ break falls just shortward of the $J$ band. At this
redshift, an $R\!-\!J$ or $I\!-\!J$ {\it vs.} $J\!-\!K$ plot provides excellent discrimination between
OGs and most other objects with similar $R\!-\!K$ values \citep{poz00,sto01}.  \citet{cim03} 
claim that some star-forming galaxies may have colors similar to those of OGs; however, 
their Fig.~5, on which
this claim is based, seems to depend strictly on {\it morphological} classes judged by eye from
{\it HST} ACS images, rather than on {\it spectroscopic} evidence for star formation.  This
approach assumes that the generally tight correlation between morphology and star-formation
rate that applies at the present epoch continues to hold at high redshifts.  One of the 
purposes of the present paper is to discuss evidence that this assumption may not always be valid.

We have used a photometric sieve process to isolate candidates.
The initial observations were carried out at $K'$ with NSFCam on the
NASA Infrared Telescope Facility (IRTF). If any objects were found within
a 30\arcsec\ square region surrounding the quasar (limited by the NSFCam field
and the dithering amplitude) and with $17\le K'\le19$ (we use the Vega
magnitude scale throughout this paper), we next obtained $J$-band imaging.
If $J\!-\!K'\approx 2$ (as expected for old stellar populations in this redshift
range), we then obtained $R$ (and sometimes $I$ and/or $z'$) imaging, generally with
the Echellette Spectrograph and Imager (ESI) on the Keck II telescope, but
sometimes with other CCD systems and telescopes. Since we were 
searching for systems of old stars with essentially no recent star formation, we
required $R\!-\!K'\ge6$; this is, of course, by itself, a necessary but not sufficient condition
at redshifts $z\sim1.5$ for the absence of young stars, because of the possible 
effects of reddening. However, together with the constraint that $J\!-\!K'\approx 2$,
this $R\!-\!K'$ color requires a sharp inflection in the spectrum consistent with the
4000 \AA\ break for old stellar populations at $z\sim1.5$.

\begin{figure}[!t]
\epsscale{0.6}
\plotone{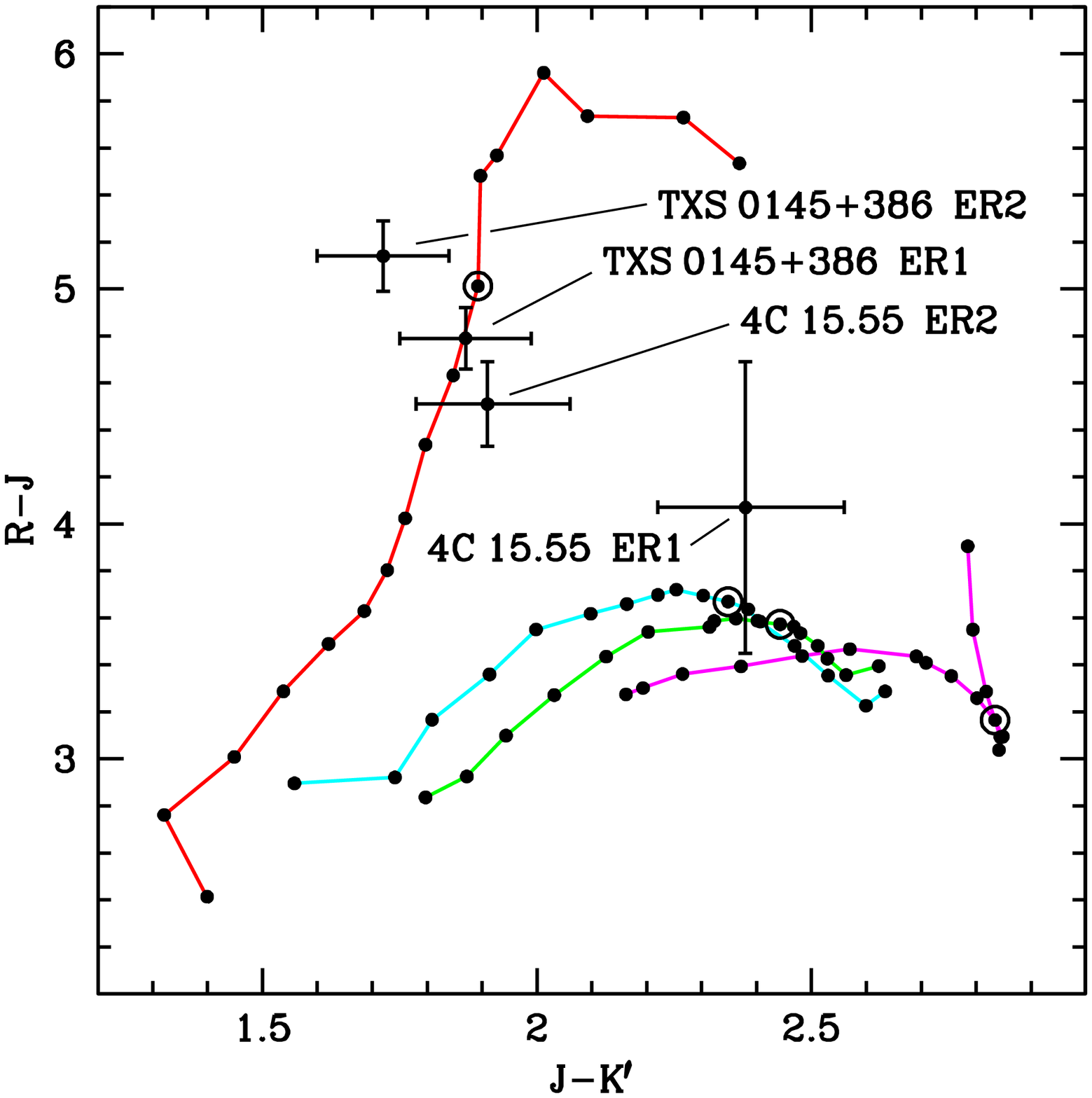}
\caption{Positions of model tracks and the objects discussed in this paper in the
$J\!-\!K'$ vs. $R\!-\!J$ plot.  Each of the model tracks ranges from $z=0.4$ to $z=2.0$
(bottom to top or left to right);
increments of 0.1 in $\Delta z$ are indicated by the black dots, with the $z=1.4$ point
circled.  The red trace is
that of a passively evolving stellar population for which all of the stars were formed
instantaneously when the universe was $5\times10^8$ years old (corresponding to
$z=9.5$). The blue trace shows a continuous star-formation model with an exponentially 
decreasing star-formation rate having an age of 12 Gyr and an $e$-folding time of 5 Gyr,
with sufficient \citet{cal00} reddening to produce $R\!-\!K'=6.0$ at $z=1.4$.
The green and magenta traces show models for starbursts, with a constant star-formation
rate and \citet{cal00} and \citet{fer99} reddening, respectively, both also normalized to
give $R\!-\!K'=6.0$ at $z=1.4$.  See 
the text for details. \label{twocolor}}
\end{figure}

We show traces for an unreddened, passively evolving galaxy and for various models of heavily
reddened galaxies as functions of redshift in the $J\!-\!K'$---$R\!-\!J$ diagram in 
Fig.~\ref{twocolor}, along
with the positions in this diagram of the 4 objects discussed in this paper.  The
passively evolving galaxy, deliberately chosen to be an extreme case, 
is assumed to have formed all of its stars $5\times10^8$
years after the Big Bang.  Reddened starbursts are represented by constant-star-formation-rate
models, with \citet{cal00} and \citet{fer99} reddening adjusted to give $R\!-\!K'=6.0$
at $z=1.4$.  The reddening curve of \citet{fer99} used here is that of a pure disk with
embedded dust and an inclination of 70\arcdeg\ to the line of sight.  Finally, we also show a model
for continuous, exponentially decreasing star formation, with an $e$-folding time
of 5 Gyr and an age of 12 Gyr, again subjected to \citet{cal00} reddening to give 
$R\!-\!K'=6.0$ at $z=1.4$.  The age of this model is of course unrealistic at the redshifts we are
considering, but it serves as a proxy for a range of models for which there
is a substantial old stellar component coupled with a relatively modest star-formation rate
and high reddening.
We will present observations of the larger sample elsewhere. We restrict ourselves here
to the \txs\ and \fourc\ fields because we have obtained more complete observations
for these fields than for any of the others, including spectroscopy of one OG in each
field with the Low-Resolution
Imaging Spectrograph (LRIS; \citealt{oke95}) on Keck I and/or ESI on Keck II. The 
spectroscopy with LRIS always used a 1\arcsec\ slit or slitmask
with a 400 line/mm grating blazed at 8500 \AA, and the ESI spectroscopy always used
a 1\farcs25 slit and the echellette mode. Other follow-up
observations on the \txs\ and \fourc\ fields include deep imaging
with the University of Hawaii 88-inch Telescope (UH88) and the United Kingdom
Infrared Telescope (UKIRT), and adaptive optics (AO) imaging with the
Canada-France-Hawaii Telescope (CFHT) and the Subaru Telescope \citep{kob00}.
For the galaxy morphology analysis from AO images, we have obtained PSF information 
from the closest sufficiently bright star in the field (12\arcsec\ away for \txs\ ER1 and
9\arcsec\ away for \fourc\ ER2). By mapping out the PSF variation with position
from other stars in the field and from other fields obtained on the same nights, we 
confirm that there should be no significant change in the PSF over these distances
on the nights of our AO observations.

A listing of all of the optical/IR observations relevant to this paper is given in Table \ref{obslog}.
\begin{deluxetable}{l l l l c c c l}
\tablenum{1}
\tablewidth{0pt}
\tabletypesize{\scriptsize}
\tablecaption{Optical and IR Observations of the TXS\,0145+386 and 4C\,15.55 Fields}
\tablehead{
\colhead{Date (UT)} & \colhead{Telescope} & \colhead{Instrument} & \colhead{Target}
& \colhead{Filter} & \colhead{Exposure} & \colhead {Pixel Scale } &
\colhead{Comments} 
}

\startdata
1998 Aug 17 & IRTF & NSFCam & 4C\,15.55 Field & \phd$K'$ & \phn\phn540 s & 0\farcs30\phn & \nodata \\
1998 Aug 17 & IRTF & NSFCam & TXS\,0145+386 Field & \phd$K'$ & \phn\phn540 s & 0\farcs30\phn & \nodata \\
1998 Aug 18 & IRTF & NSFCam & 4C\,15.55 Field & $J$ & \phn\phn540 s & 0\farcs30\phn & \nodata \\
1998 Aug 18 & IRTF & NSFCam & TXS\,0145+386 Field & $J$ & \phn\phn540 s & 0\farcs30\phn & \nodata \\
1998 Aug 19 & IRTF & NSFCam & 4C\,15.55 Field & \phd$K'$ & \phn\phn540 s & 0\farcs30\phn & \nodata \\
1998 Aug 19 & IRTF & NSFCam & 4C\,15.55 Field & $J$ & \phn1620 s & 0\farcs30\phn & \nodata \\
1998 Aug 19 & IRTF & NSFCam & TXS\,0145+386 Field & \phd$K'$ & \phn1080 s & 0\farcs30\phn & \nodata \\
1998 Aug 19 & IRTF & NSFCam & TXS\,0145+386 Field & $J$ & \phn1080 s & 0\farcs30\phn & \nodata \\
1998 Sep 05 & UH88 & QUIRC & TXS\,0145+386 Field & $J$ & \phn3000 s & 0\farcs18\phn & \nodata \\
1998 Oct 28 & UH88 & Tek2048 & TXS\,0145+386 Field & \phd$R_c$ & 12000 s & 0\farcs22\phn & \nodata \\
1998 Oct 29 & UH88 & Tek2048 & TXS\,0145+386 Field & \phd$I_c$ & 14400 s & 0\farcs22\phn & \nodata \\
1999 Aug 24 & CFHT & Pueo/KIR & TXS\,0145+386 Field & \phd$K'$ & \phn6000 s & 0\farcs035 & \nodata \\
1999 Oct 02 & Keck I & LRIS & TXS\,0145+386 Field & \phd$Z$ & 16200 s & 0\farcs22\phn & \nodata \\
1999 Nov 03 & Keck I & LRIS & TXS\,0145+386 ER1 & Spec & 16800 s & 0\farcs22\phn & 1\arcsec\ slit \\
2000 Jun 05 & Keck II & ESI & 4C\,15.55 ER2 & Spec & 18000 s & 0\farcs15\phn & 1\farcs25\phn slit, echellette mode \\
2000 Jun 05 & Keck II & ESI & 4C\,15.55 ER1 & Spec & \phn3600 s & 0\farcs15\phn & 1\farcs25 slit, echellette mode \\
2000 Oct 03 & Keck II & ESI & TXS\,0145+386 ER1 & Spec & 21600 s & 0\farcs15\phn & 1\farcs25 slit, echellette mode \\
2000 Oct 04 & Keck II & ESI & TXS\,0145+386 ER1 & Spec & 19120 s & 0\farcs15\phn & 1\farcs25 slit, echellette mode \\
2001 Aug 24 & Keck I & LRIS & TXS\,0145+386 ER1 & Spec & 10800 s & 0\farcs22\phn & slitmask \\
2002 May 31 & Subaru & CISCO & 4C\,15.55 Field & \phd$K'$ &\phn1920 s & 0\farcs11\phn & \nodata \\
2002 May 31 & Subaru & CISCO & 4C\,15.55 Field & $J$ & \phn\phn960 s & 0\farcs11\phn & \nodata \\
2002 Aug 07 & Keck II & ESI & 4C\,15.55 Field & \phd$R_c$ & \phn2520 s & 0\farcs15\phn & Ellis R filter \\
2002 Aug 17 & Subaru & IRCS/AO & 4C\,15.55 Field & \phd$K'$ & \phn2160 s & 0\farcs058 & \nodata \\
2003 Dec 05 & UKIRT & UFTI & TXS\,0145+386 Field & $K$ & \phn8400 s & 0\farcs090 & \nodata \\
\enddata
\label{obslog}
\end{deluxetable}
The optical and IR imaging observations were processed through our standard IRAF reduction
pipelines.  The spectroscopic data needed more detailed and individual attention, both in the
planning of the observations and the reduction, because of the faintness of the objects. For the
longslit observations, both with LRIS and with ESI, we arranged a slit position angle that would
allow a brighter object to be placed near one end of the slit while the actual target was more-or-less
centered. After the individual spectra were flat-fielded, distortion corrected, and wavelength
calibrated, the spectrum of the bright object could be used to register the spectra for coadding before
extracting the spectrum of the target object.  For the ESI observations, the bright object 
was also necessary to confirm that the echellette orders were properly normalized to each other
in the overlap regions before they were combined to make a continuous spectrum.

\subsection{X-Ray Observations}

The purpose of the Chandra X-ray observations was to search for evidence
for cluster gas surrounding the two quasars.
The TXS\,0145+386 and 4C\,15.55 fields were observed with ACIS-I 
on Chandra
on January 4, 2002 UT for 63 ksec, and on April 13--14, 2002 UT for
52 ksec, respectively. ACIS-I consists of four CCDs with a total
field of view of $16\arcmin\times16\arcmin$, covering an energy range from
0.3 to 10 keV.  At full resolution, the plate scale is 0\farcs5 per
pixel. The images were obtained in the VFAINT mode, which potentially offers
better rejection of spurious events and therefore provides a lower background,
at the possible expense of exacerbating pileup problems near bright sources.  
We analyzed both the pipeline-reduced images, which used the 
standard algorithm for grading events, and images we reprocessed
using the VFAINT algorithm.

Using the {\it Chandra Interactive Analysis of Observations} (CIAO)
software\footnotemark[3]\footnotetext[3]{http://cxc.harvard.edu/ciao/}, we produced exposure
maps for each observation which we then used to flat-field the images
and convert to flux units.  In order to simplify the calculation of the
exposure map, we assumed a monoenergetic distribution of source photons
for both observations.  Ideally one would prefer to define a set of
spectral weights for the image in order to more accurately determine the
exposure map, which is a complicated function of energy.  However, given
that each of the sources are strongly peaked at energies of $\sim1.1$--1.3 keV,
and that exposure maps produced at slightly higher and lower
energies produced negligible difference in the final fluxed images, we
are confident that the error introduced by using the monoenergetic approximation is small.

For each observation, source and background regions were chosen from the
final fluxed image and then applied to the raw counts image.  Several
background regions were chosen for statistical robustness, and the dispersion
in the background regions was found to be approximately 17\% for
TXS\,0145+386 and 14\% for 4C\,15.55; these percentages are similar to the calculated
Poisson errors.  From the counts image, we calculated the Poisson upper
and lower confidence limits for the background and source regions using
the values given by \citet{geh86} and \citet{ebe03}.  

We computed model PSFs for each source using the Chanda Ray
Tracer (ChaRT) tool, which contains detailed information about the High
Resolution Mirror Assembly (HRMA) on Chandra, and allows for highly
accurate determinations of the PSF over a range of energies and off-axis
positions.  Using CIAO's modeling and fitting program, Sherpa, we were
able to model the spectrum of each source with a 1-D power law combined
with photoelectric absorption.
This spectrum was passed to ChaRT in order to determine the ray traces,
which were then projected onto the ACIS-I detector plane, and converted to
a FITS image using MARX.  Both TXS\,0145+386 and 4C\,15.55 are bright
sources and suffered from photon pileup; however, MARX allowed us to
simulate the PSF both with and without pileup.  We find that while pileup
affects the innermost pixels in the PSF, the outer wings of the
PSF ($>5$\arcsec) are essentially unaffected, even with the VFAINT processing,
so that an accurate subtraction of
the contribution from the PSF can still be obtained in the region relevant to our
search for cluster emission.

\section{Results}

\subsection{The TXS\,0145+386 Field}

\begin{figure}[!tbh]
\epsscale{0.8}
\plotone{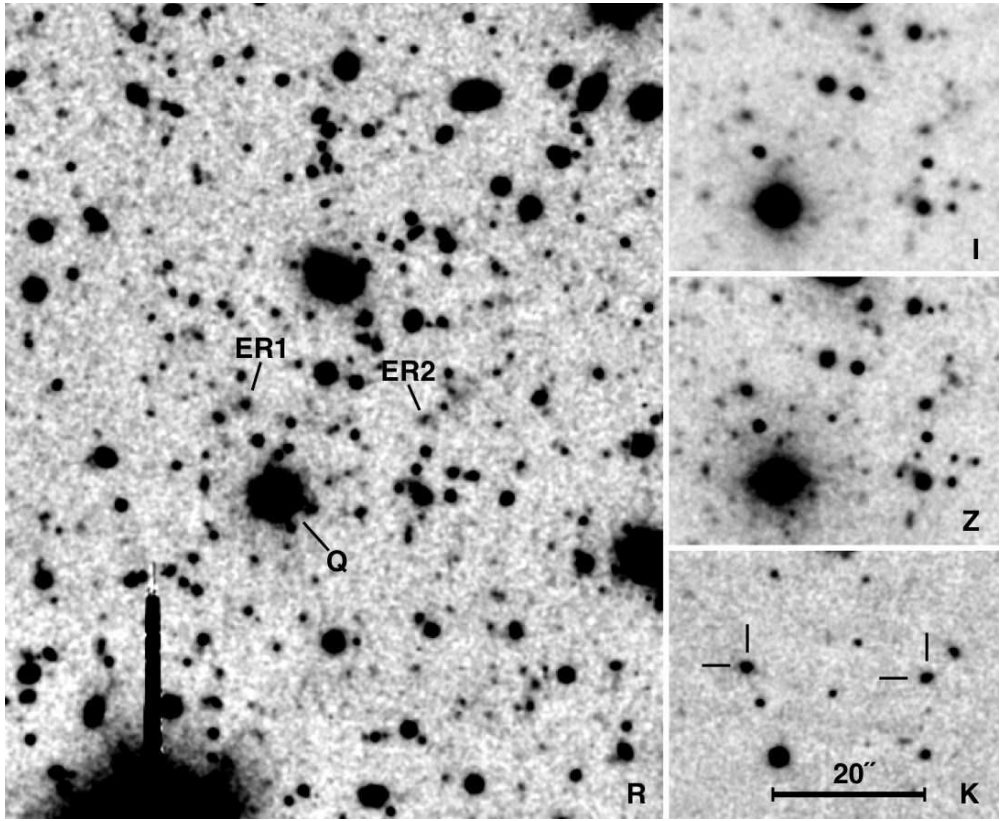}
\figcaption{Images of the TXS\,0145+386 field. The quasar TXS\,0145+386 itself is 
indicated by the ``Q''.  The two EROs are labelled ``ER1'' and ``ER2'' and are marked
by the short lines in the $K$ image.\label{txsimage}}
\end{figure}

\begin{figure}[!t]
\epsscale{0.6}
\plotone{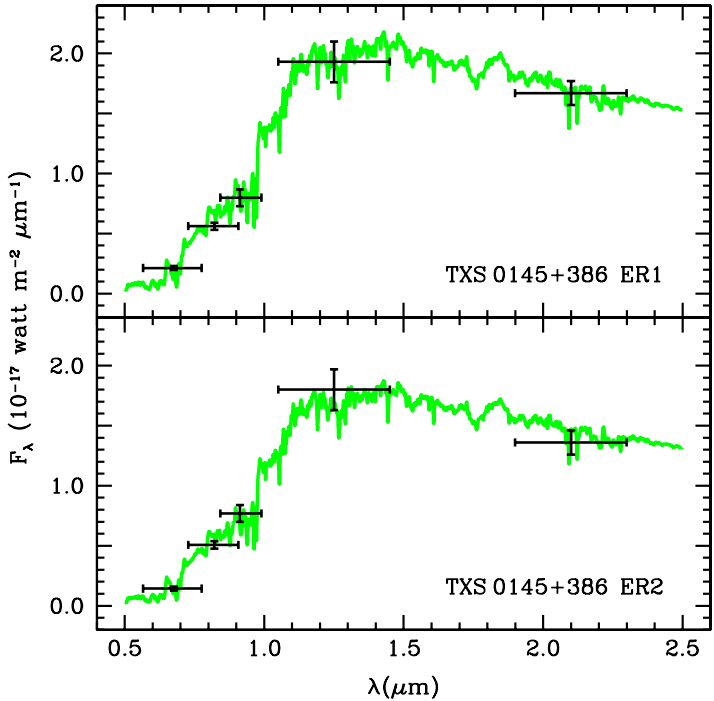}
\figcaption{Photometry of two EROs found in the TXS\,0145+386 field. The photometric
data are shown in black, with the vertical bars indicating $1 \sigma$ uncertainties and the 
horizontal bars indicating filter half-transmission widths. The green curves show a
Bruzual \& Charlot (2003) 3-Gyr-old instantaneous burst model for reference.\label{txsphot}}
\end{figure}

\begin{figure}[p]
\epsscale{0.5}
\plotone{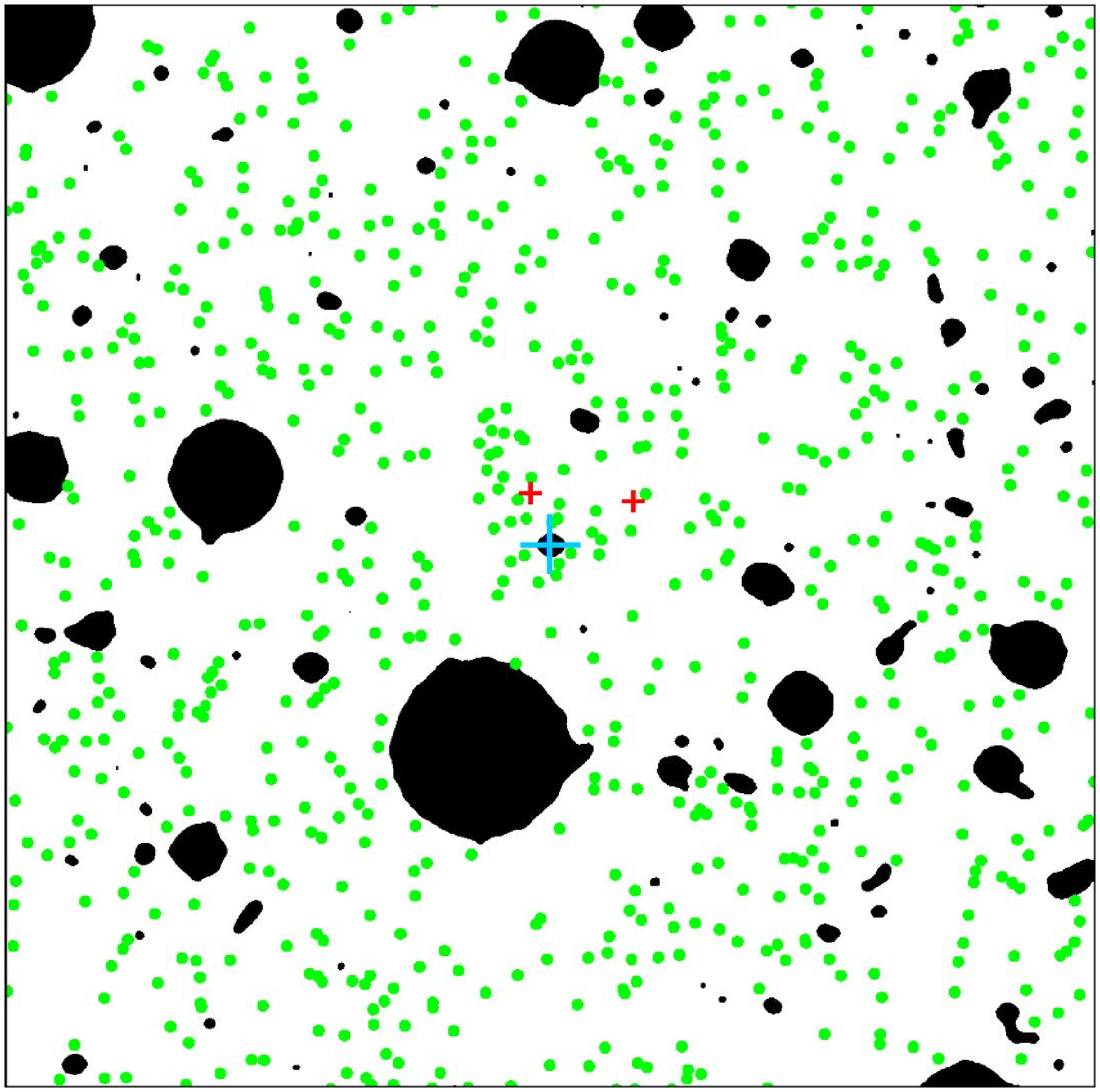}
\figcaption{Evidence for clustering of faint galaxies in the vicinity of TXS\,0145+386.
The green dots show the distribution of galaxies with $24\le z\le25.5$ in a $4\farcm3\times4\farcm3$
field around TXS\,0145+386. Black areas are regions obscured by bright stars or 
galaxies.  The two EROs are indicated by red crosses, and the quasar by a blue cross.
\label{txsclus}}
\end{figure}

Images in several bandpasses of the \txs\ field are shown in Fig.~\ref{txsimage}. Two 
galaxies that meet our selection criteria are found in this field; these are labelled ``ER1''
and ``ER2'' in the figure. The spectral-energy distributions (SEDs) for these galaxies
are shown in Fig.~\ref{txsphot}, along with a reference spectrum for a 3-Gyr-old
solar-metallicity instantaneous-burst model \citep{bru03}.
There appears to be an excess of faint objects in Fig.~\ref{txsimage} surrounding the
quasar and ER1. This impression is confirmed by Fig.~\ref{txsclus}, which shows a
4\farcm3-square region centered on \txs, with positions of all galaxies in a magnitude
interval $24\le z\le25.5$ indicated.
\begin{figure}
\epsscale{0.8}
\plotone{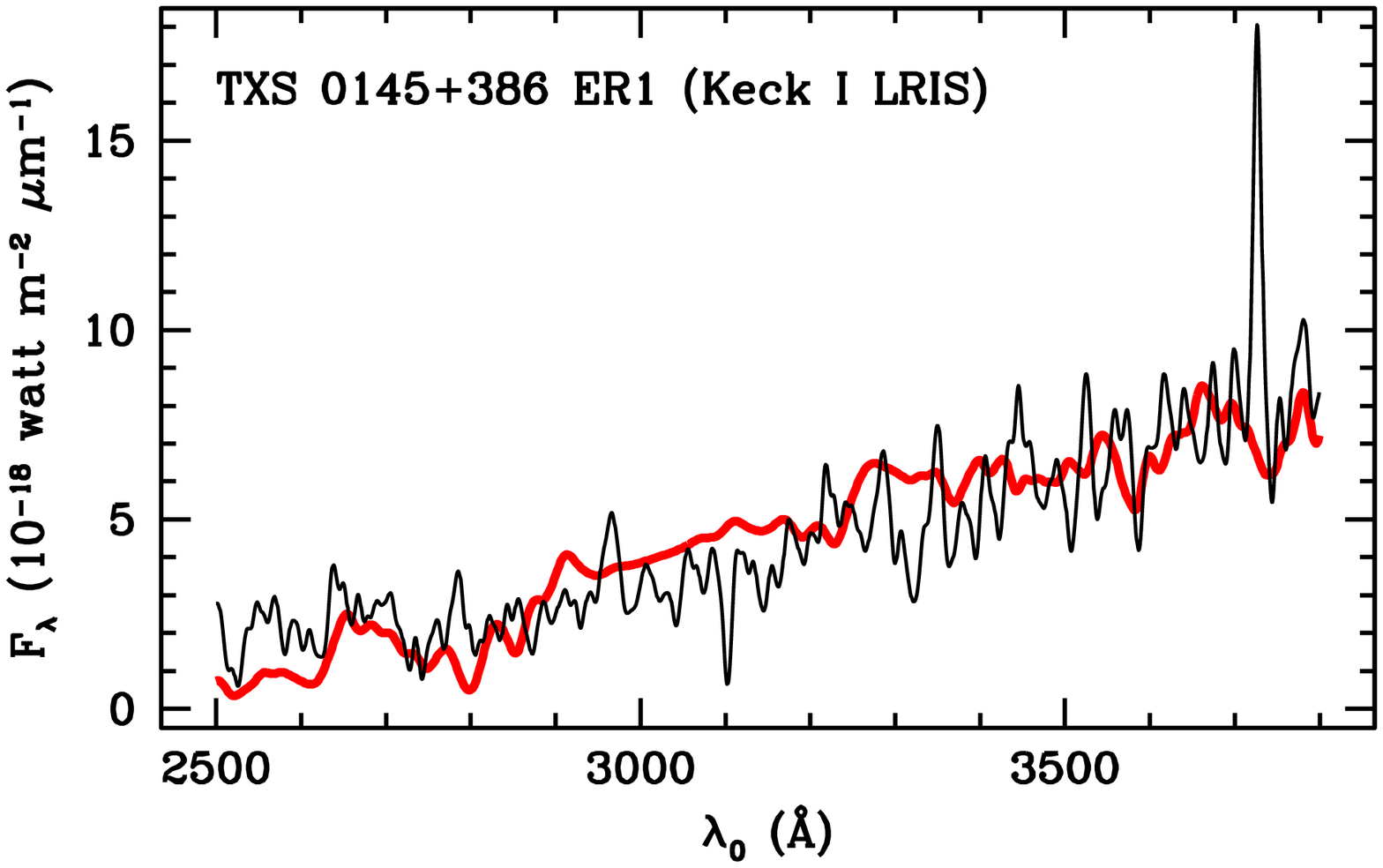}
\figcaption{The spectrum of TXS\,0145+386 ER1 ({\it black line}), obtained with LRIS. 
The red trace shows
the Bruzual \& Charlot (2003) 3-Gyr instantaneous-burst model. Note the [\ion{O}{2}]
$\lambda3727$ emission in the galaxy spectrum, which gives a redshift 
$z=1.4533$ for the galaxy.\label{txsspec}}
\end{figure}
We have attempted to obtain spectroscopy of \txs\ ER1. Figure \ref{txsspec} shows
that the spectrum is generally consistent with that expected for an old stellar population
at the approximate redshift of the quasar.  However, since the S/N ratio for this spectrum
is not very high and because there are no strong absorption 
features in the available spectral range, it is also probably consistent with the average ``dusty
star-forming'' spectrum of \citet[][see their Fig.~2]{cim02}. The presence of significant star formation
would seem, at least at first sight, to be corroborated by the presence of 
[\ion{O}{2}] $\lambda3727$ emission. However, spectra obtained at
higher resolution with ESI on Keck II show that the emission has an intrinsic velocity
width of $900$ km s$^{-1}$, considerably higher than would be expected simply from ordered
rotation or other gravitationally induced motions of star-forming regions in even the most 
massive galaxies. Such
velocities could indicate the presence of wind, driven either by a hidden starburst or
by a hidden AGN. On the other hand, as we describe 
in more detail below, there is no evidence for a point X-ray
source that would indicate the presence of an AGN at this location in our deep Chandra ACIS images.
The absence of obvious \ion{Mg}{2} or [\ion{Ne}{5}] emission also favors stellar excitation over
an AGN origin.
If the emission is due to recent star formation, the [\ion{O}{2}] luminosity indicates a 
current star-formation rate (uncorrected for internal extinction) of $\sim2$ $M_{\odot}$ yr$^{-1}$, 
using the conversion given for high-redshift galaxies by \citet[][eq.~19]{kew04}. This rate is
very similar to the rate found by \citet{cim02} for their sample of dusty star-forming EROs.

In sum, the [\ion{O}{2}] emission seems most reasonably interpreted as indicating a moderate
degree of star formation (depending, of course, on the amount by which the [\ion{O}{2}]
emission is affected by internal extinction); the large velocity width remains a puzzle.  However,
the SED (Fig.~\ref{txsphot}), with its strong break at redshifted 4000 \AA, clearly
matches that of an old stellar population. We know of no way to
reproduce a similar SED with any reasonable reddening of a stellar population 
whose rest-frame UV-optical luminosity is dominated by young stars.  It seems safe
to conclude that the observed-frame $K'$-band morphology of \txs\ ER1 is
overwhelmingly dominated by light from old stars.

\begin{figure}[!tbh]
\epsscale{0.6}
\plotone{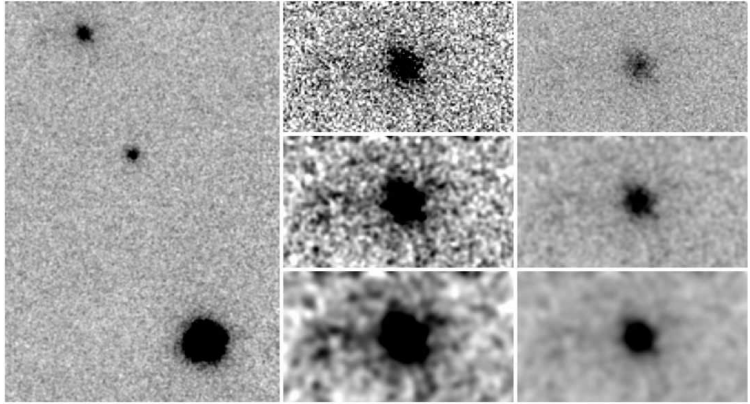}
\figcaption{AO imaging of \txs\ ER1 with PUEO and KIR on CFHT.  In the left panel, 
the quasar is at the
bottom, ER1 at the top.  The object near the center is an extremely compact galaxy shown
by our spectroscopy to be at $z=0.8$.  The right panels show, from top to bottom, two
different contrast stretches for each of three degrees of smoothing of enlarged images
of \txs\ ER1.  Although there are clearly problems with uneven background levels
in the higher contrast images, there seems to be some evidence for 
irregularities in the low-surface-brightness outer parts
of ER1.\label{txsao}}
\end{figure}

\begin{figure}[!tbh]
\epsscale{0.5}
\plotone{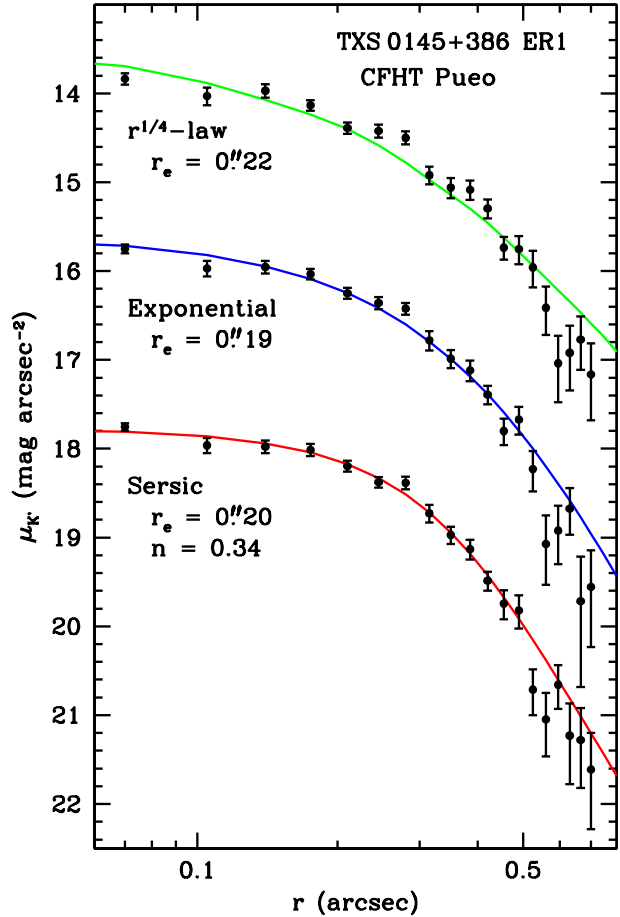}
\figcaption{Seeing-convolved radial surface-brightness plots fitted to the observed
radial-surface-brightness profile of \txs\ ER1.  The surface-brightness scale to the
left is correct for the bottom (red) curve and points; the other curves and their
corresponding points are displaced by 2 and 4 magnitudes. 
The compact nuclear component has
been removed from both the models and the data.  Because the model centers for
both the galaxy and the nucleus vary slightly among the various models, as does
the ellipticity distribution (as a function of radius), we have used these model centers 
and ellipticities to constrain the determination of the observed profiles in each case.
\label{txs0145sb}}
\end{figure}

There is a mag 12 star about 1\arcmin\ south of \txs\ ER1. Although this offset is rather
larger than one would like, we were able to obtain AO imaging with PUEO on CFHT 
on a night during which the angular coherence was exceptionally good, as indicated
by the round PSFs of the quasar and the nearly stellar galaxy between the quasar and ER1.
The $K'$ AO image is shown in Fig.~\ref{txsao}.  \txs\ ER1 itself shows an elliptical 
image with a high degree of symmetry, surrounded by some irregular structure at very low 
surface brightness.
Determining the radial surface-brightness profile for \txs\ ER1 from the AO image is
complicated by the fact that there is an unresolved core present, which is offset
from the centroid of the underlying galaxy by $\sim0\farcs12$ ($=1$ kpc).  Because
both the relative centroids and the amplitudes of the core and galaxy components
vary slightly in different models, we have carried out independent modeling of the
galaxy in terms of $r^{1/4}$-law, exponential, and S\'{e}rsic profiles, using {\sc galfit}
\citep{pen02}.  In each case, we included a PSF component as well to represent
the star-like core, and we allowed 
the centers of the galaxy and PSF components to float freely.  Figure \ref{txs0145sb}
shows fits of PSF-convolved radial-surface-brightness plots for each of the models
to the measured profile, after removal of the compact nucleus.
The best-fit S\'{e}rsic model has a quite low index of 0.34.
Even so, the points at large radius still tend systematically to fall below the curve.  
The exponential model is nearly as good a fit, and the $r^{1/4}$-law model
is clearly worse, being too high at small and large radii and too low at
intermediate radii.  For the S\'{e}rsic model, the compact core comprises about
30\% of the total light in the $K'$ band.

We do not have AO imaging of \txs\ ER2; nevertheless, our best $K'$ images
show a somewhat asymmetric structure suggesting that this galaxy may be
an ongoing or recent merger.  If so, the fact that the SED is clearly that of
an old stellar population with no hint of recent star formation indicates that
the participating galaxies are likely to have been largely free of cold gas.

For the X-ray statistics, we chose to use an inner radius of 10\arcsec\ and
an outer radius of 30\arcsec\ for our annulus around TXS\,0145+386.  The
outer limit was chosen at this scale because of a nearby X-ray source that
would have confused the statistics had we used a larger radius.  This
outer radius corresponds to a physical scale of 253 kpc at $z=1.442$.
After subtracting background counts and the modeled wings of the PSF
in the annulus, we are left with an insignificant 3 (+12, $-$9) excess counts.
In order to estimate what this result means in terms of limits on cluster X-ray
emission, we assume a specific reasonable model:  a modified King profile (so-called $\beta$
model) with $\beta=0.7$, a core radius of 140 kpc, a metallicity of 0.4 solar, and
a temperature $kT=6$ keV (cf. \citealt{mul05}), and centered on the quasar. 
Taking all of the flux out to 10 core radii, we find that the annulus contains
24.5\%\ of this total, so the extrapolated total residual counts would be
12 (+49, $-$37).  We use WebPIMMS\footnotemark[4]
\footnotetext[4]{http://heasarc.gsfc.nasa.gov/Tools/w3pimms.html}
to calculate the total luminosity corresponding to a $3~\sigma$ upper limit of
approximately 150 counts ($2.4\times10^{-3}$ counts s$^{-1}$), assuming a
Raymond-Smith model for the SED. This 
limit comes out to $5.7\times10^{44}$ ergs s$^{-1}$ for a rest-frame energy range
from 0.5--8 keV.  This limit is not a very strong constraint; for example, the cluster
recently found by \citet{mul05} at a similar redshift has $\sim3\times10^{44}$ erg s$^{-1}$
over the 0.5--2 keV range, which, given the difference in energy range, 
is similar to our upper limit.

There is no significant evidence for X-ray emission from either \txs\ ER1
or \txs\ ER2.  Formally, for ER1, we have $3\pm3\times10^{-5}$ counts per second, 
and, for ER2, $0\pm3\times10^{-5}$ counts per second.  In contrast, for
\txs\ itself, we have $1.75\pm0.05\times10^{-2}$ counts per second.

\subsection{The 4C\,15.55 Field}

\begin{figure}[p]
\epsscale{0.8}
\plotone{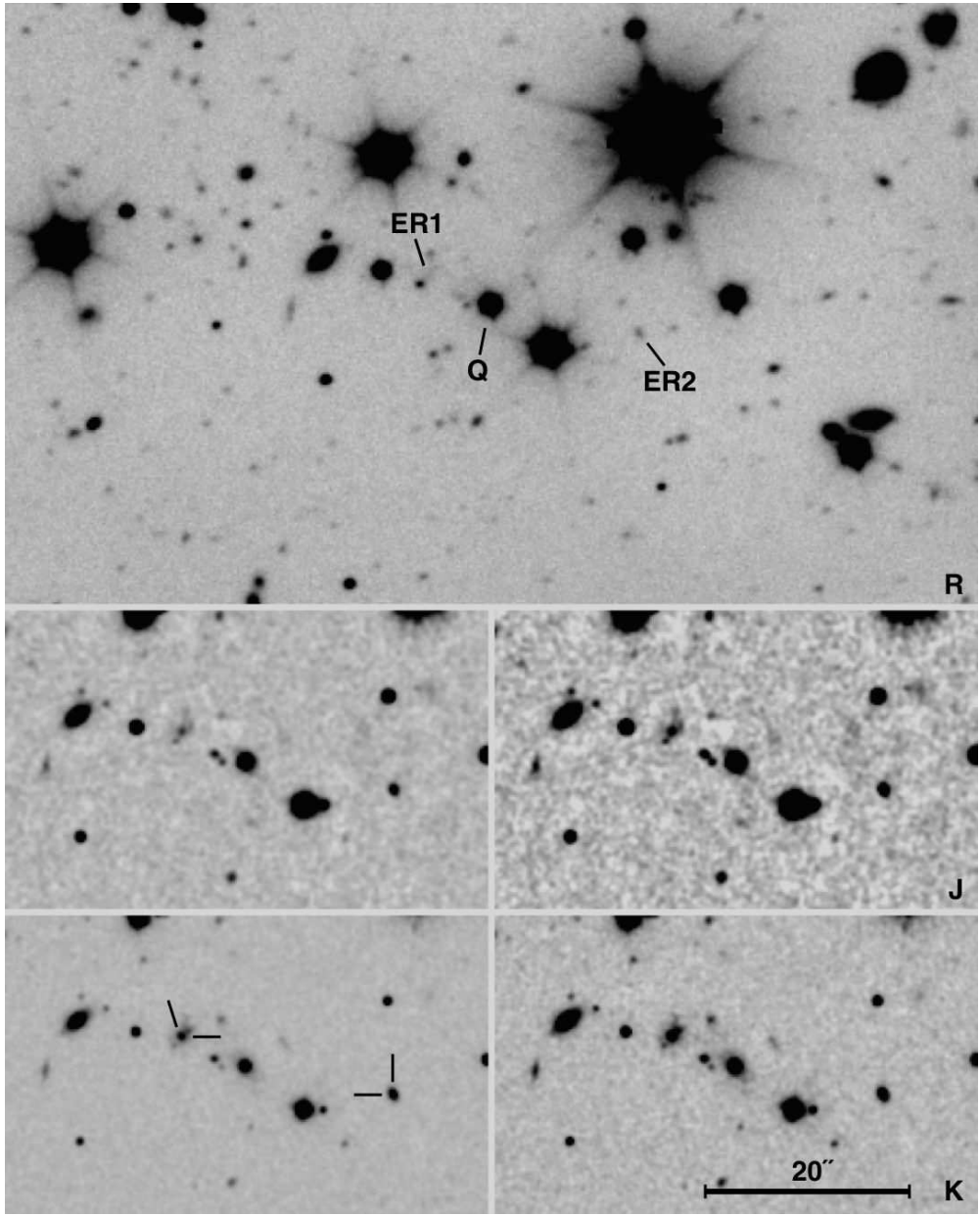}
\figcaption{Images of the 4C\,15.55 field. The quasar 4C\,15.55 itself is 
indicated by the ``Q'', and the two EROs are labelled ``ER1'' and ``ER2''. Note the difference in morphology between ER1, which the photometry indicates is likely
a dusty star-forming galaxy, and ER2, which clearly has the characteristics of an old stellar 
population with little reddening (see Fig.~\ref{4cphot}). The left and right versions
of the $J$ and $K'$ images show different contrast levels.\label{4c15mos}}
\end{figure}

\begin{figure}[!b]
\epsscale{0.6}
\plotone{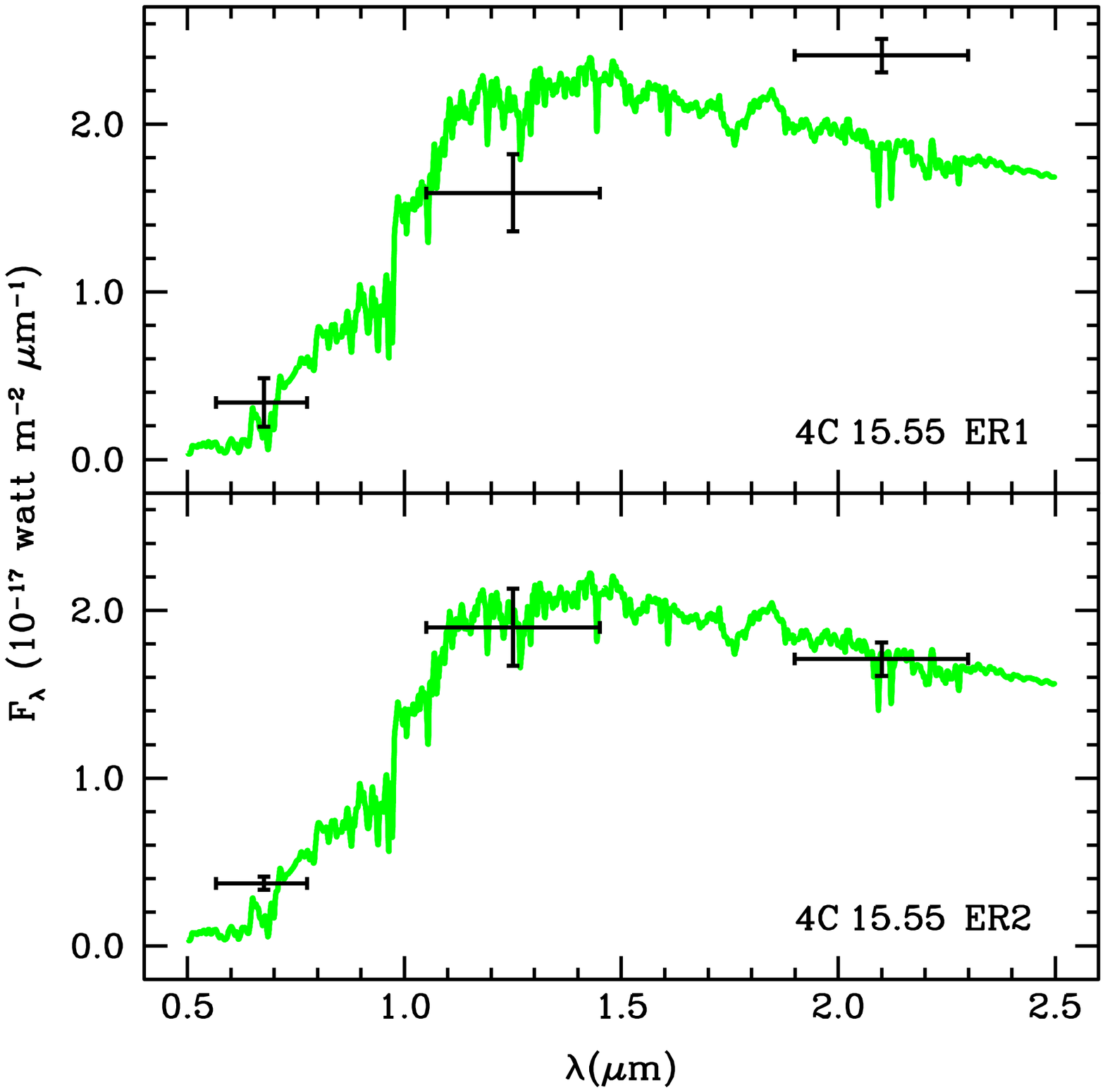}
\figcaption{Photometry of two EROs found in the 4C\,15.55 field. The photometric
data are shown in black, with the vertical bars indicating $1 \sigma$ uncertainties and the 
horizontal bars indicating filter half-transmission points. The solid green curves show
scaled Bruzual \& Charlot (2003) 3-Gyr-old solar-metallicity instantaneous burst models for 
reference.\label{4cphot}}
\end{figure}

\begin{figure}[!tbh]
\epsscale{0.8}
\plotone{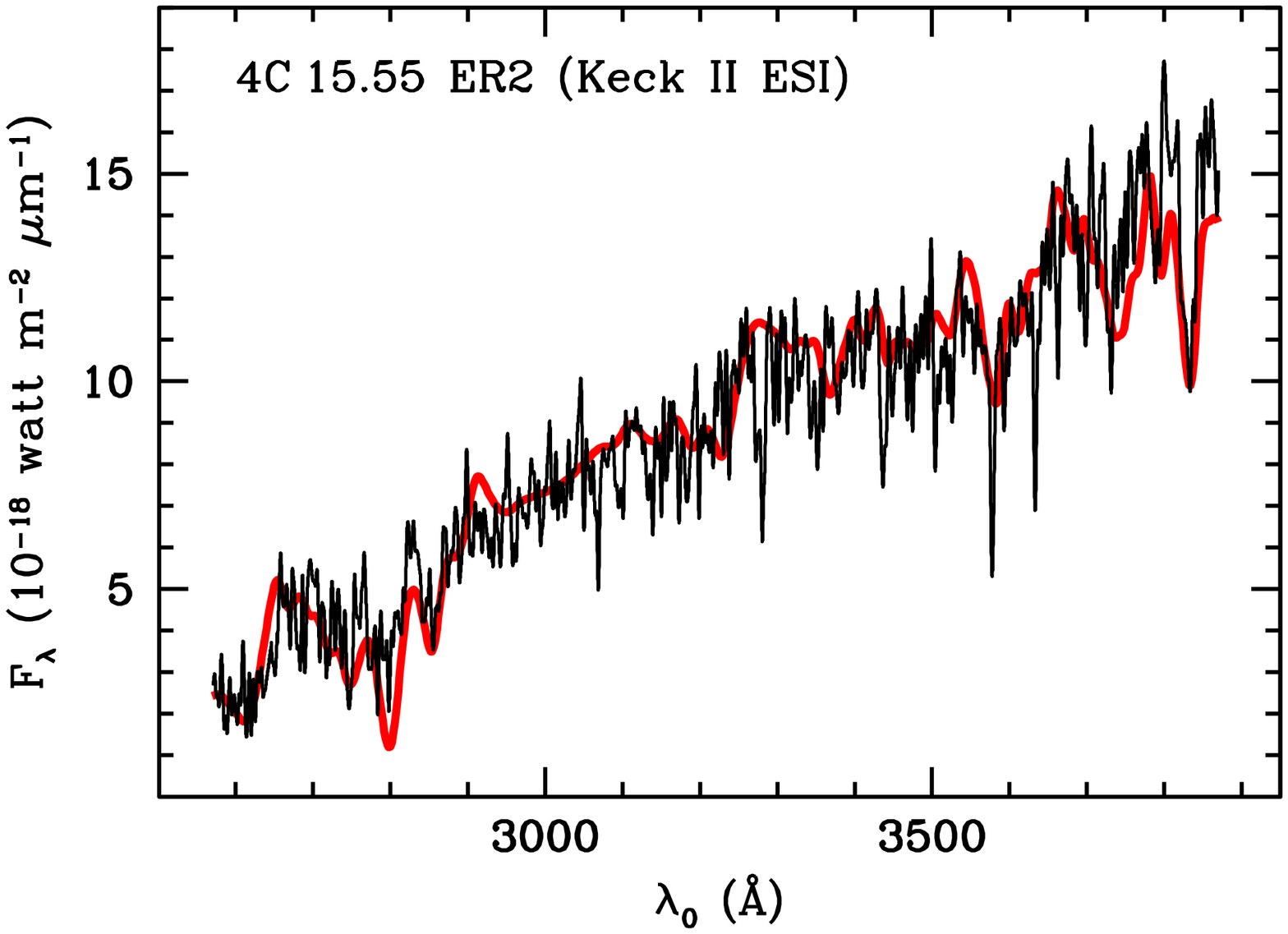}
\figcaption{The spectrum of 4C\,15.55 ER2 ({\it black line}). The red trace shows
the Bruzual \& Charlot (2003) 3-Gyr 0.32 solar-metallicity instantaneous-burst model.
The redshift of 4C\,15.55 ER2 is 1.412.\label{4cspec}}
\end{figure}

\begin{figure}[!tbh]
\epsscale{0.6}
\plotone{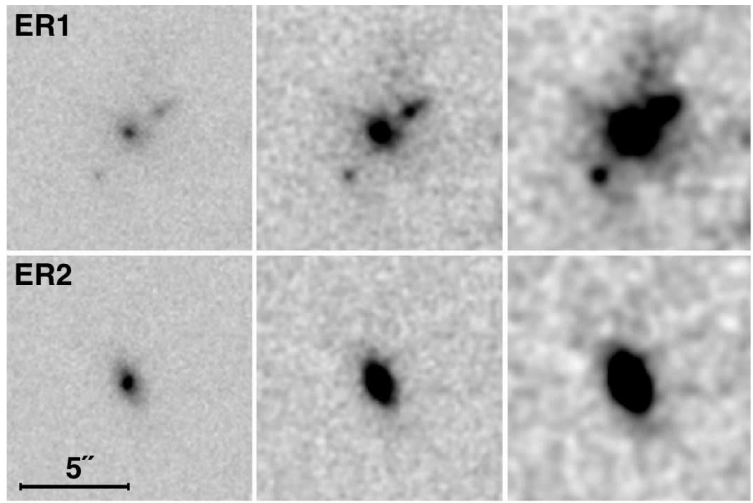}
\figcaption{Images of 4C\,15.55 ER1 and ER2 obtained with the Subaru AO system and
IRCS. The left-hand panels show the images smoothed with a Gaussian kernel with
$\sigma=0.5$ pixel at low contrast; the middle and right-hand panels are at successively
higher contrasts smoothed with $\sigma=1.0$ pixel and $\sigma=2.0$ pixels, respectively.\label{4cao}}
\end{figure}

\begin{figure}[p]
\epsscale{0.6}
\plotone{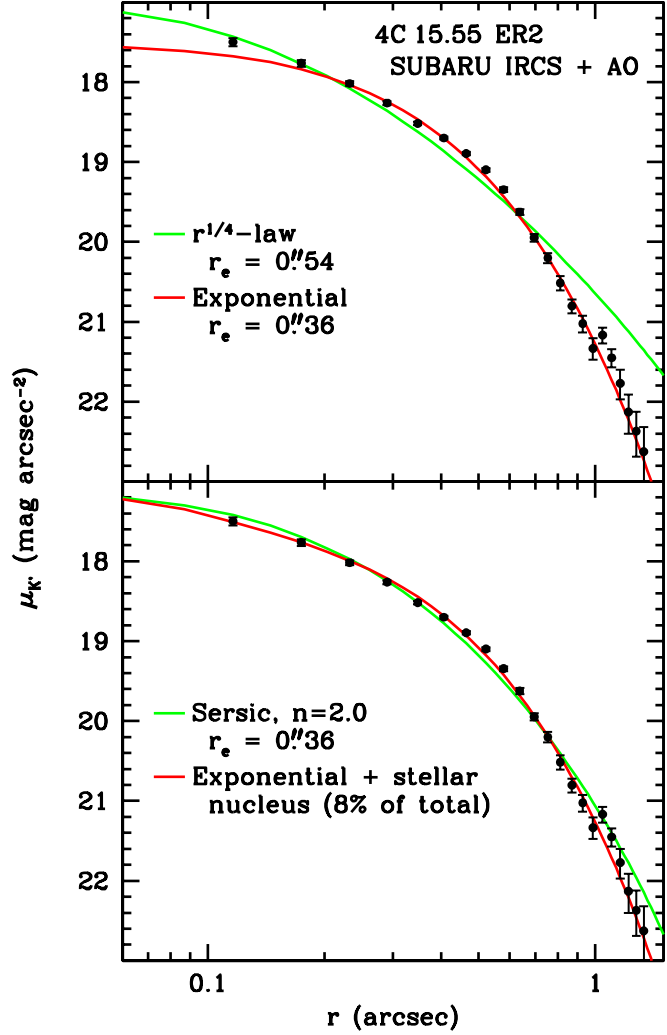}
\figcaption{The surface-brightness profile of 4C\,15.55 ER2, from AO imaging
obtained with the Subaru AO system and IRCS. The colored curves show
fits of various functions to the data, all convolved with the point-spread function. 
The green curve in the upper panel shows an $r^{1/4}$-law fit, normalized
to the central data points, but falling below them at intermediate radii and
above them at large radii. The red curve shows an exponential profile with
$r_e=0\farcs36$, which is an excellent fit
beyond 0\farcs2, but under-fits the central region. In the lower panel, the
green curve shows the best-fit S\'{e}rsic profile, with $n=2.0$, and the red curve
shows a best-fit composite profile, adding an 8\%\ contribution from a stellar or
compact nucleus to the exponential profile.\label{4csbp}}
\end{figure}

\begin{figure}[!t]
\epsscale{0.6}
\plotone{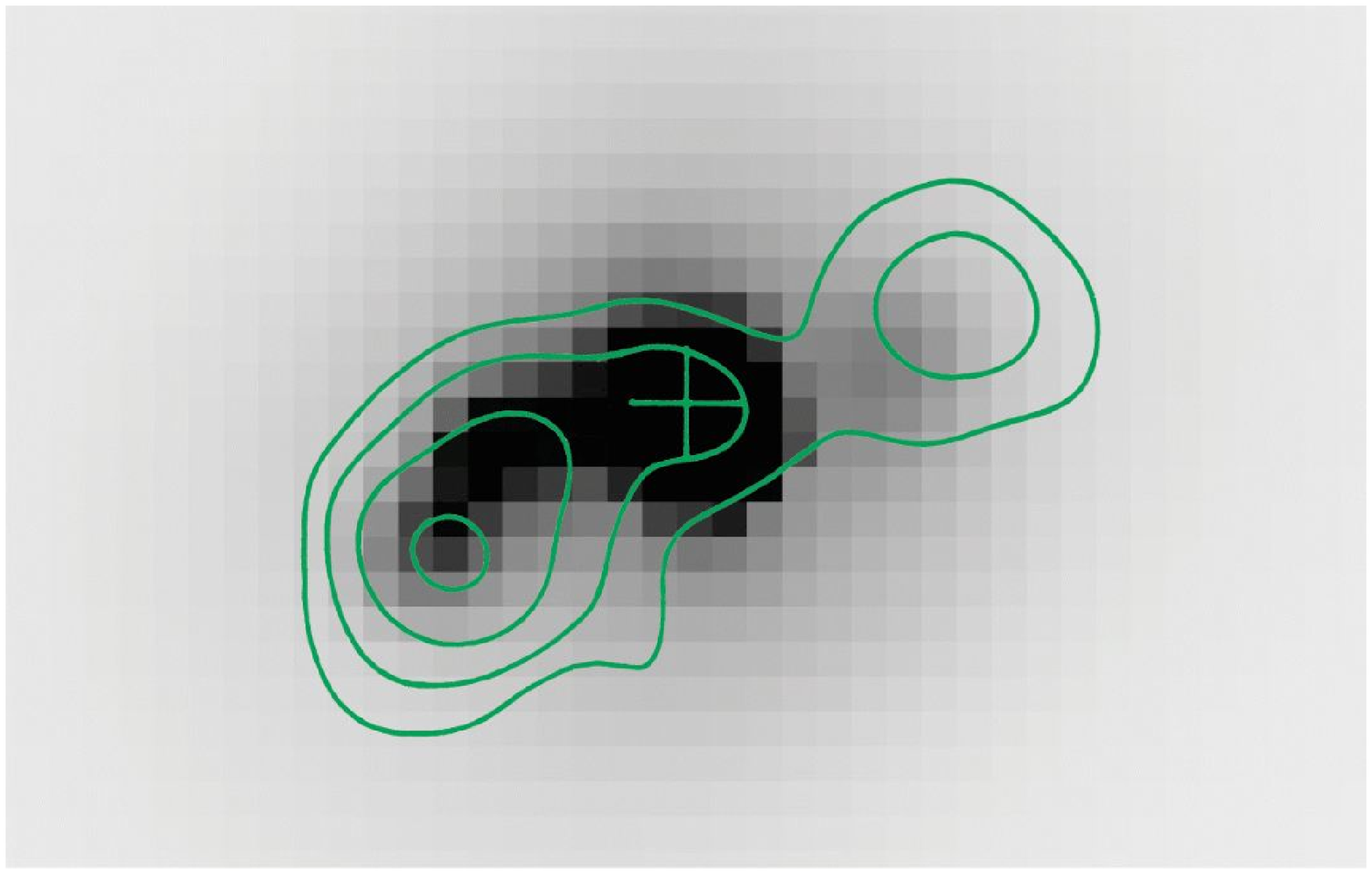}
\figcaption{The X-ray jet of 4C\,15.55.  The gray-scale ACIS image has been
adaptively smoothed with the CIAO {\it csmooth} routine, and the green contours 
are 20 cm radio emission data from the
VLA A-array observations of \citet{hin83}. The cross marking the central radio component
has been centered on the peak of the X-ray emission. The scale can be judged from the
pixels, which are 0\farcs5 square.\label{4c15jet}}
\end{figure}

The field of the quasar 4C\,15.55 is shown in Fig.~\ref{4c15mos}. Again, there are
two EROs in the field. However, our photometry (Fig.~\ref{4cphot}) indicates that
ER1, though quite red ($R\!-\!K'\sim 6.5$), does not show the strong break characteristic
of an old stellar population (see also Fig.~\ref{twocolor}); we therefore classify it as a likely 
dusty star-forming galaxy. On the
other hand, ER2 in this field clearly meets our criterion for an old stellar population.
The rather large uncertainty shown for the $R$-band photometry of ER1 in Fig.~\ref{4cphot} 
results from having to correct for the effect of the blue object about 2\arcsec\ south-east of
ER1.
Figure \ref{4cspec} shows our spectrum of \fourc\ ER2, obtained with ESI on Keck II.
The S/N of this spectrum is sufficiently high that we can see clear matches between
absorption features in the spectrum and those in the 3-Gyr \citet{bru03} 
model. The galaxy has a redshift of 1.412, slightly higher than that of the quasar,
which is given as 1.410 by \citet{smi77} and as 1.406 by \citet{wil78}.
Our AO imaging of the \fourc\ field is shown in Fig.~\ref{4cao}.  In order to have
a large enough field to image both EROs in this field, we used the 0\farcs058
pixel scale option. ER1, which we had inferred on photometric grounds to be a reddened 
star-forming galaxy, is clearly an interacting system, while ER2 shows a very 
regular, symmetric profile. 
\clearpage
We have carried
out two-dimensional, PSF-convolved modeling of the profile of ER2, using {\sc galfit} and
the STSDAS {\it ellipse} task.  As Fig.~\ref{4csbp} shows, a pure $r^{1/4}$-law is a poor
fit to the profile. An exponential model provides an excellent fit beyond a semi-major axial radius
of about 0\farcs2 but falls below the observed profile at the center. A S\'{e}rsic profile
with $n=2.0$ comes reasonably close to fitting the points at all radii, although it is
slightly above at both large and small radii and slightly below in the middle.
Our best fit is achieved with 
a composite profile, comprising a dominant exponential with the addition of a small stellar
component to the nucleus (this ``stellar nucleus'' could be a $r^{1/4}$-law
bulge or a nuclear disk with a small effective radius; in any case, the contribution 
to the total light from this compact central component is small).
For the Chandra ACIS data for 4C\,15.55, the analysis is made difficult by the 
presence of an X-ray jet associated with the QSO (see Fig.~\ref{4c15jet}).  
This jet likely contributes to the extended emission
around the QSO, but, while the contribution from the central PSF can be
modeled, the jet's full contribution cannot be known precisely.  We
therefore tried to avoid any region where jet emission would confuse the
statistics.  We chose a circular aperture of radius 50\arcsec\ around the
QSO (corresponding to $\sim$700 kpc at $z=1.5$), excluding an elliptical
region with a semi-major axis also of 50\arcsec\ and a semi-minor axis of
15\arcsec\, and with a position angle of $-68\fdg2$, corresponding to the observed
jet position angle.  We measured the flux
in background regions using the same aperture and calculated the Poisson
errors.
After accounting for the contribution from the quasar's PSF and subtracting
the background, we find an excess of $37\pm16$ counts around the source.  We
note that the interpretation of this detection remains somewhat ambiguous, 
given the presence of 
the jet, but that an overdensity of events is clearly visible around the source in the
original Chandra image.
On the assumption that this overdensity can be attributed to an extended cluster
component rather than diffuse radiation associated with the X-ray jet, we estimate
the X-ray luminosity of the cluster using the same assumptions as we did for obtaining 
an upper limit for the TXS\,0145+386 field. In this case, extrapolating the counts in our
aperture to 10 core radii gives a total of $93\pm39$. Again using WebPIMMS, with the
same parameters as before, we find a total luminosity of 
$4.0\pm1.7\times10^{44}$ ergs s$^{-1}$. This is again similar to the luminosity found
by \citet{mul05} for their non-AGN-related cluster at the slightly lower redshift of 1.39.

As for \txs\ ER1 and ER2, \fourc\ ER2 itself shows no evidence for point-like
X-ray emission.

\section{Discussion}

The clustering amplitude of luminous galaxies comprising old stellar populations at high
redshifts \citep{dad00,mcC01,bro05} is similar to that of early-type galaxies at the present epoch.
It is also comparable to that of powerful radio sources at similarly high redshifts, so
it would not be surprising if these two classes were to have a considerable overlap in
their spatial distribution.  \citet{hal01} and \citet{wol03} have found an excess of 
EROs in radio-loud quasar fields that tends to corroborate such an overlap.  \citet{wol03}
suggested two possibilities for this excess: (1) a physical clustering of early-type galaxies
around the quasars, and (2) massive clusters along the line of sight to the quasars, which
bias the quasar catalogs via gravitational amplification.  Our spectroscopic confirmation
that at least one of the OGs in each of our fields has a redshift agreeing with that of the
quasar to within a few hundred \kms\ supports the first interpretation.

Our attempt to detect direct evidence for clusters in the two fields we discuss in this paper
has produced ambiguous results.  While we clearly seem to find an enhanced
surface density of faint galaxies in the field of TXS\,0145+386, our Chandra ACIS 
observation of the same field gives only an upper limit to any diffuse X-ray emission
from the cluster.  On the other hand, in the field of 4C\,15.55, we appear to have a
$\sim2.4~\sigma$ X-ray detection of diffuse emission, but we cannot be absolutely sure 
that this emission
is not associated with the X-ray jet from the quasar, in spite of our effort
to excise any such emission. Neither the upper limit in the first case, nor the possible detection
in the second, are inconsistent with the presence of a fairly massive cluster
surrounding the respective quasars. 

Both of the OGs for which we have AO imaging appear to have compact
nuclear components. 
After removal of these nuclear components, both of these galaxies
are well fit by exponential profiles (and not 
by $r^{1/4}$-law profiles); however, the discrimination among profiles is much better for
\fourc\ ER2 because of the much higher S/N of the Subaru image.  For this galaxy,
there is no significant doubt that the stellar distribution has close to an exponential
profile; even if one were to prefer the $n=2$ S\'{e}rsic model (which eliminates the need
to add a compact nucleus), it is only slightly different from an exponential
beyond a radius of 0\farcs2 ($=1.7$ kpc), and, to the extent it differs, it is a worse
fit to the data.  It is true (as the referee has emphasized) that elliptical galaxies
in the present-day universe can have a wide range of S\'{e}rsic indices, 
including some as low as 2 \citep{caon93,kho00}.  However, the same studies
that have demonstrated this range have also shown that there is a strong positive correlation 
between the S\'{e}rsic index and both the central surface brightness and the total 
luminosity of the galaxy.
Ellipticals with $n\sim2$ are, virtually without exception, galaxies with luminosities
considerably below $L^*$ and with low
central surface brightnesses (see, \eg\ Table 2 of \citealt{caon93}). There are
good reasons to believe that they have had different formation histories than have
luminous ellipticals \citep{kho00}, possibly involving secular evolution from disks.
Such galaxies are in any case
quite different from \fourc\ ER2, which is luminous ($\sim2L^*$) and has a high surface
brightness. 

Some massive galaxies strongly dominated by old stellar populations at high 
redshifts clearly show closely $r^{1/4}$-law profiles \citep{bun03,cim04,fu05,dad05}; others
just as clearly show essentially pure exponential profiles \citep{sto04,fu05,dad05}.  The only 
reasonably large structures
in the present-day universe that show exponential profiles are the disk components
of spiral and S0 galaxies and the ``pseudo-bulges'' \citep[\eg][]{kor04} found in late-type 
spirals.  The latter are not very massive, and they are produced over a significant fraction
of a Hubble time via secular evolution from disks.  For these reasons,
they are almost certainly not models for the exponential profiles we see in high-redshift OGs.
This means that the exponential profiles in these massive high-redshift systems
are due either to disks or to some configuration that is not found among massive
galaxies at the present epoch. In at least one case \citep{sto04}, the galaxy has a 
small axial ratio and looks very much like a highly inclined disk.

If these really are massive ($\sim2\times10^{11} M_{\odot}$) disks of old stars, then they
are significantly smaller than most massive disks at the present epoch. For example,
if we take from the study of a complete sample of nearby disks by \citet{vdK87} those
with disk $M<-20.2$ (for our assumed $H_0=70$ km s$^{-1}$ Mpc$^{-1}$), the average
exponential scale length is $5.5\pm2.7$ kpc, which corresponds to an effective
radius $r_e=9.2\pm4.5$ kpc. In comparison, \txs\ ER1 and \fourc\ ER2 have 
$r_e = 1.69$ kpc and 3.0 kpc, respectively. Even the larger of these is barely
more than half as large as the smallest of the 12 luminous galaxies from the 
\citet{vdK87} sample. \citet{dad05} have similarly found that early-type galaxies
at high redshifts are significantly smaller than their counterparts in the present-day
universe. Eight out of their 9 examples were found to have morphologies
consistent with spheroids, and only 1 appeared to be a disk of old stars. But it is perhaps
significant that this disk-like galaxy, at $z=2.47$, was by a fair margin the highest redshift
galaxy in their sample.  \citet{dad05} point out that this galaxy is quite similar to the 
galaxy discussed by \citet{sto04}, which is also at a similar redshift.

Such massive disks would
carry certain implications for the early formation of massive galaxies:
\begin{enumerate}
\item Such a disk must have formed quasi-monolithically and with strong dissipation. It is
unlikely to have suffered any major mergers after a substantial fraction of the stars had 
been formed.
\item Depending on the age of the stellar population in the examples observed so far,
quite large and sustained star formation rates are required to form most of the stellar
mass in a short time. At a minimum, star formation rates of a few hundred $M_{\odot}$
per year sustained for a few $10^8$ years are indicated.  Note that similar constraints are 
indicated by the $\alpha$-element enhancement of the stellar content of massive
ellipticals at the present epoch (\eg\ \citealt{wor92, wor98}).
\item Since massive pure disks of old stars do not exist (or at least are extremely rare) 
at the present epoch, such 
galaxies must have a very high probability of being transformed into ellipticals via
major mergers or into bulge-dominated S0s via less extreme encounters. This would not be surprising,
given that our objects have been found in radio-source fields, which should be among 
the highest density regions in the early universe.  At high redshifts, most of these regions
will not yet have become virialized, so velocity dispersions in sub-clumps may well be
conducive to merging activity.  The fact that we suspect TXS\,0145+386 ER2 to be an
ongoing merger gives some observational credence to this line of reasoning.  This
scenario might also explain why, among the admittedly small number of OGs that so far
have published high-resolution morphological data, spheroids seem to be more common at
$z\sim1.5$ than at $z\sim2.5$.
\end{enumerate}

All of this fits in rather well with what we know or suspect about massive cluster 
ellipticals at the present epoch.  Numerous lines of evidence \citep[\eg][]{pee02,tho02,tho05} 
point to early and
rapid formation of the great majority of the stars in the most massive and luminous
ellipticals, yet some simulations \citep[\eg][]{tho99,kau00} suggest late 
($z$ between 1 and 2) assembly of the galaxies
themselves.  The only way these two conditions can be reconciled is if the assembly
is done from galaxies that formed most of their stars at much earlier times and that
are largely free of gas.  Furthermore, the assembly process must be dominated by (at
most) a few large such galaxies in order to preserve the well-known color-luminosity relation
for ellipticals.  Thus, early forming massive disks in which star formation has been driven to
completion or truncated by the expulsion of remaining gas could plausibly act 
as reservoirs for the 
stars that eventually end up in present-day elliptical or cD galaxies.

\acknowledgments
We thank the anonymous referee, whose prompt and detailed report helped
us improve both the presentation and content of the paper.
This research has been partially supported by NSF grant AST03-07335
and by the National Aeronautics and Space Administration through Chandra 
Award Number GO2-3180X 
issued by the Chandra X-ray Observatory Center, which is operated by the 
Smithsonian Astrophysical Observatory for and on behalf of the National 
Aeronautics Space Administration under contract NAS8-03060. 
It made use of the NASA/IPAC Extragalactic Database (NED) 
which is operated by the Jet Propulsion Laboratory, California Institute of 
Technology, under contract with the National Aeronautics and Space 
Administration. The authors recognize the very significant
cultural role that the summit of Mauna Kea has within the indigenous
Hawaiian community and are grateful to have had the opportunity to
conduct observations from it.

\clearpage

\end{document}